\documentclass[11pt,twoside]{article}
\usepackage{asp2014}

\aspSuppressVolSlug
\resetcounters

\begin{document}

\title{TOPCAT Visualisation over the Web}

\author{Mark~Taylor}
\affil{H.~H.~Wills Physics Laboratory, Tyndall Avenue,
       University of Bristol, UK;
       \email{m.b.taylor@bristol.ac.uk}}

\paperauthor{Mark~Taylor}{m.b.taylor@bristol.ac.uk}{0000-0002-4209-1479}{University of Bristol}{School of Physics}{Bristol}{Bristol}{BS8 1TL}{U.K.}



  
\begin{abstract}

The desktop GUI catalogue analysis tool TOPCAT,
and its command-line counterpart STILTS,
offer among other capabilities visual exploration
of locally stored tables containing millions of rows or more.
They offer many variations on the theme of scatter plots,
density maps and histograms, which can be navigated interactively.
These capabilities have now been extended to a client-server model,
so that a plot server can be run close to the data storage,
and remote lightweight HTML/JavaScript clients can configure
and interact with plots based on that data.
The interaction can include pan/zoom/rotate navigation,
identifying individual points, and potentially subset selection.
Since only the pixels and not the row data are transmitted to
the client, this enables flexible remote visual exploration of
large tables at relatively low bandwidth.
The web client can request any of the plot options available
from TOPCAT/STILTS.
Possible applications include web-based visualisations
of static datasets too large to transmit,
visual previews of archive search results,
service-configured arrays of plots for complex datasets,
and embedding visualisations of local or remote tables
into Jupyter notebooks.
  
\end{abstract}

\section{Introduction}

TOPCAT \citep{2005ASPC..347...29T} is a desktop application for
manipulating tabular data, typically source catalogues.
It offers many features, intended to enable astronomers to
use their skills for understanding the scientific meaning in
tables rather than wrangling numbers.

Its visualisation windows in particular offer excellent
facilities for interactive exploration of multi-dimensional
tabular data with high row and column counts;
locally stored multi-million-row datasets can be explored with ease,
even on modestly specified laptop or desktop machines.
Many different 2D and 3D plot types and options are available,
and the display is equally capable of revealing
the overall structure of a large dataset or the individual
characteristics of a single source \citep{informatics4030018}.

In recent years, much application software has become available
as web applications;
a good example is Aladin-Lite \citep{2017ASPC..512..105B}
and various products built around it.
This mode of software provision offers various
advantages including convenience for the user and central control
of updates and configuration.
The TOPCAT client application as a whole is not very suitable for running
within a web browser for a number of reasons,
including the complexity of the GUI
(which typically requires use of many separate windows during a session)
and requirements for local filesystem access
(memory mapping is used extensively for efficient data access,
but is not permitted within a browser sandbox).
It is however both feasible and useful to provide TOPCAT's
facilities for interactive visualisation within a web browser,
and this paper explains how this has been done.

\section{Remote Visualisation Architecture}

\begin{figure}
\includegraphics[width=\textwidth]{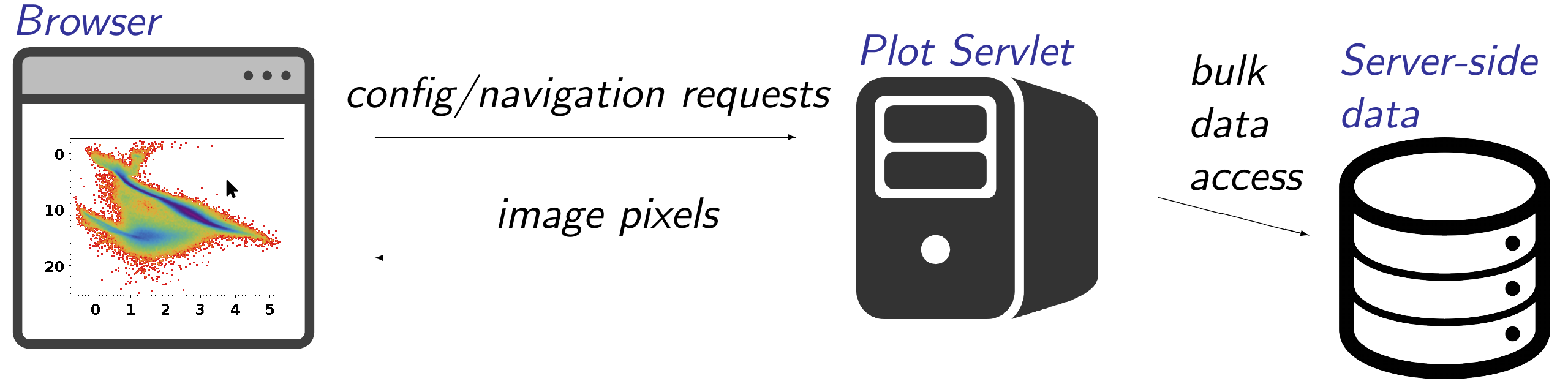}
\caption{Client-server architecture for remote visualisation.}
\end{figure}

The interactive visualisation within TOPCAT and its command-line
counterpart STILTS is handled by an internal library with the
informal name {\em Plot2}.
This library handles all plotting operations apart from the
user interaction itself; in particular it reports what
plot types and options are available;
renders an image given a suitable set of values for such options,
and updates that image as required based on supplied user
navigation actions such as mouse wheel and drag gestures.
The library also supplies other services such as
conversion between graphics and data coordinates,
reporting data bounds of the current plot, and
identifying table row indices in a given region of graphics space.
TOPCAT and STILTS each provides its own user interface (UI) layer
to talk to this backend visualisation service.

This is a somewhat simplified outline of the actual
UI-backend interaction, which also involves considerable complexity
related to session management and data caching,
but it captures two important facts:
\begin{enumerate}
\item The UI layer requires only an abstract view
of the nature of a particular visualisation, and in particular
does not need to be aware of the (many) different
plot types and configuration options on offer to the user;
it simply queries the library for the options,
gets the user to fill in the values, and passes those values
back to the library in order to acquire a display image.
This means that the complexity of the UI layer is not coupled
to the number of different options available,
so that adding new plot types or configurations requires no change
to the UI code.
\item The plot information handled by the UI layer
is in the form of images
(typically pixel arrays, though vector graphic representations
can optionally be used) rather than plotted points.
This means that the required bandwidth between the UI and the backend
does not scale with the number of table rows or plotted points.
\end{enumerate}
The separation of concerns described in (1)
makes it relatively straightforward
to implement additional UI layers, so a web-based front end has
recently been added.  This consists of a {\em Plot2\/}-based Java servlet
that can run on the server side close to the data to be visualised,
intended to serve requests from relatively thin JavaScript clients
hosted in web pages or applications.
A supplied client JavaScript library a few hundred lines long
provides a straightforward API to set up interactive plots using the same
{\em name}={\em value} configuration syntax used when plotting from STILTS.
All the plot types and configuration options are available,
and the powerful STILTS expression language can be used for
column manipulation and row selection.

Because of (2), the client-server communication is restricted to download of
image data; since table data is never exchanged over the wire,
client and network resource usage remains modest for plots of
even very large server-side tables.
This ``dumb client'' approach contrasts with the ``smart client''
implementation used by a number of general purpose JavaScript
plotting libraries which download all the data from the server
when a plot is specified, and subsequently perform all rendering
within the browser.  Such smart client plotting can provide
more responsive plots for modest-sized data sets, but becomes
unworkable because of download times and browser memory usage
with point counts in the region of 10$^{5}$ or 10$^{6}$.

\section{Usage}
\label{O1-22:usage}

The interactive plots embedded in web pages by this service
can be zoomed, panned, and in the case of 3D plots rotated,
using familiar mouse gestures exactly as in TOPCAT's plot windows.
They can also be configured to interact with the input row data,
for instance to display the full row content relating to a point that
the user clicks on.
There are various ways that this client-server visualisation
could be employed.

One scenario is for data providers or archives
to deliver a quick look at available data for users before download.
An archive might provide one or more ``front-page'' interactive views
of the content of a survey
(feasible for a few million or maybe tens of millions of rows),
or provide on archive query result pages the option to display
source positions, colour-magnitude diagrams, or user-specified plots,
which an archive user could investigate interactively before,
or instead of, downloading the result table.

Another option would be for a scientist to provide on a personal
web page some interactive plots related to recent research results.
In this case complex visualisations could be set up that allow readers to
investigate the data in more detail than can be done with the
static figures of a published paper.

The HTML embedding an interactive plot can also be inserted
into the output cells of a Jupyter notebook,
so that a Jupyter user can create and configure visualisations
programmatically by issuing python commands.
In the context of a Science Platform this could allow
users to interact visually with data sets they have assembled
on a remote server without ever having to download it locally.

Finally, a more ambitious client-side possibility would be to
write a JavaScript application that allows the user to set up and
configure such visualisations graphically rather than preconfigure
them in static or dynamic web pages.

\section{Deployment}

The visualisation server implementation is in the form of a java Servlet.
It can therefore be installed in a running servlet container such as Tomcat,
but it can also be run standalone directly from the STILTS application
(by executing ``{\tt stilts server}''),
which contains an embedded instance of the Jetty servlet engine.
A Tomcat-based Docker image has also been prepared for convenience.

In all these cases the data files to be visualised
have to be made available to the servlet;
the arrangements for doing this
vary slightly for the different deployment options,
but essentially the tables just need to be in storage near to the server.
Any STIL-friendly storage format is suitable,
though FITS,
or for large/wide tables column-oriented FITS \citep{2008ASPC..394..422T},
is recommended; no other special preparation of the table data is required.
It should also be possible to plot SQL database query results directly
using JDBC, though at time of writing this has not been tested.

The servlet also requires some temporary storage space for caching
prepared coordinate data and images to improve performance when
updating plots during visualisation sequences.
On the server side more processing cores, faster access to data storage,
more memory,
and more and faster scratch space will improve performance;
servlet configuration options are available to manage usage
of these resources.

Resource usage on the client side is minimal,
apart from bandwidth for image data download,
which as noted above, does not scale with table row count.
Typically, and depending on server load and network characteristics,
a few frames per second can be delivered to the user.
While this falls short of gaming-quality animation,
it is sufficient in most cases for interactive data exploration.

\section{Status and Future Work}

At time of writing support for these interactive plots is included
in the most recent release of STILTS (v3.3)
and therefore available for public use,
though it is still somewhat experimental.
The functionality is known to work,
but has not been tested on large servers or with heavy multi-user loads.
Various improvements to the implementation as it stands could
be desirable:
tuning of the caching arrangements and server-side resource usage,
attending more carefully to data access and security restrictions,
improving session management by storing more state on the client side,
adding facilities like interactive region selection,
and improvements to the supplied JavaScript client library
for instance allowing interactive configuration of plot
characteristics like colour maps as well as interactive navigation.

It is not clear which if any of the scenarios described in
Section~\ref{O1-22:usage}, or others,
will be attractive to users in practice.
Future development of these capabilities will be guided by
input from deployers or prospective deployers.
Interested individuals or projects
are encouraged to contact the author to discuss their
requirements or experience.

\bibliography{O1-22}

\begin{thebibliography}{}
\expandafter\ifx\csname natexlab\endcsname\relax\def\natexlab#1{#1}\fi
\expandafter\ifx\csname url\endcsname\relax
  \def\url#1{\texttt{#1}}\fi
\expandafter\ifx\csname urlprefix\endcsname\relax\def\urlprefix{URL }\fi
\providecommand{\eprint}[2][]{\url{#2}}

\bibitem[{{Boch} \& {Fernique}(2017)}]{2017ASPC..512..105B}
{Boch}, T., \& {Fernique}, P. 2017, in ADASS XXV, edited by N.~P.~F. {Lorente},
  K.~{Shortridge}, \& R.~{Wayth}, vol. 512 of ASP Conf. Ser., 105

\bibitem[{{Taylor}(2005)}]{2005ASPC..347...29T}
{Taylor}, M.~B. 2005, in ADASS XIV, edited by P.~{Shopbell}, M.~{Britton}, \&
  R.~{Ebert}, vol. 347 of ASP Conf. Ser., 29

\bibitem[{Taylor(2017)}]{informatics4030018}
Taylor, M.~B. 2017, Informatics, 4. \eprint{1707.02160},
  \urlprefix\url{http://www.mdpi.com/2227-9709/4/3/18}

\bibitem[{{Taylor} \& {Page}(2008)}]{2008ASPC..394..422T}
{Taylor}, M.~B., \& {Page}, C.~G. 2008, in ADASS XVII, edited by R.~W.
  {Argyle}, P.~S. {Bunclark}, \& J.~R. {Lewis}, vol. 394 of ASP Conf. Ser., 422

\end{thebibliography}


\end{document}